\documentclass[a4paper,12pt]{article}
\usepackage{amssymb}
\usepackage{amsmath,color}
\usepackage{bbold}
\usepackage{slashed}
\usepackage{epsfig}
\usepackage{graphicx,epsf,epsfig}
\usepackage{bbm}
\usepackage{subfigure}
\usepackage{multirow}
\usepackage{setspace}
\usepackage{ulem}
\usepackage{pifont}
\usepackage{authblk}
\usepackage{cite}

\newcommand{\mee}{\langle m_{ee}\rangle}

\newcommand{\obb}{0\nu\beta\beta}

\newcommand{\mwl}{m_{W_L}}
\newcommand{\mwr}{m_{W_R}}
\newcommand{\lamfunc}{\lambda\left(s,\mwl^2,\mwr^2\right)}

\newcommand{\ba}{\begin{array}{c}}
\newcommand{\baz}{\begin{array}{cc}}
\newcommand{\bad}{\begin{array}{ccc}}
\newcommand{\bav}{\begin{array}{cccc}}
\newcommand{\baf}{\begin{array}{ccccc}}
\newcommand{\ea}{\end{array}}
\newcommand{\ta}[1]{#1\hspace{-.42em}/\hspace{-0.7em}}

\def\ra{\rightarrow}
\def\be{\begin{equation}}
\def\ee{\end{equation}}
\def\gs{\mathrel{
   \rlap{\raise 0.511ex \hbox{$>$}}{\lower 0.511ex \hbox{$\sim$}}}}
\def\ls{\mathrel{
   \rlap{\raise 0.511ex \hbox{$<$}}{\lower 0.511ex \hbox{$\sim$}}}}

\newcommand{\bea}{\begin{equation} \begin{array}{c}}
\newcommand{\eea}{ \end{array} \end{equation}}

\newcommand{\csection}[1]{\section*{\centering #1}}
 
\def\slc#1{\setbox0=\hbox{$#1$}           
    \dimen0=\wd0                                 
    \setbox1=\hbox{/} \dimen1=\wd1               
    \ifdim\dimen0>\dimen1                        
       \rlap{\hbox to \dimen0{\hfil/\hfil}}      
       #1                                        
    \else                                        
       \rlap{\hbox to \dimen1{\hfil$#1$\hfil}}   
       /                                         
    \fi}

\title{
\hfill {\small IFIC/12-19}\\[0.4in]
\vskip 0.24cm
 \Large 
\bf Linear Collider Test of a Neutrinoless Double Beta Decay Mechanism 
in left-right Symmetric Theories}
\author{James Barry$^1$\footnote{E-mail: {\tt james.barry@mpi-hd.mpg.de}} }
\author{~Luis Dorame$^{1,2}$\footnote{E-mail: {\tt dorame@ific.uv.es}} }
\author{~Werner Rodejohann$^1$\footnote{E-mail: {\tt werner.rodejohann@mpi-hd.mpg.de}}}
\affil{$^1$Max-Planck-Institut f\"{u}r Kernphysik, \\ Saupfercheckweg 1, 69117 Heidelberg, Germany}
\affil{$^2$AHEP Group, Institut de F\'{i}sica Corpuscular -- C.S.I.C./Universitat de Val\`{e}ncia
Edificio Institutos de Paterna, Apt 22085, E-46071 Valencia, Spain}
\date{}
\begin{document}

\maketitle

\begin{abstract}
\noindent
There are various diagrams leading to neutrinoless double beta decay
in left-right symmetric theories based on the gauge group
$SU(2)_L \times SU(2)_R$. All can in principle be tested at a linear
collider running in electron-electron mode. 
We argue that the so-called $\lambda$-diagram is the most promising one. 
Taking the current limit on this diagram from double beta decay
experiments, we evaluate the relevant cross section $e^- e^- 
\to W^-_L W^-_R$, where $W^-_L$ is the Standard Model $W$-boson and
$W^-_R$ the one from $SU(2)_R$. It is observable if the life-time of
double beta decay and the mass of the $W_R$ are close to current
limits. 
Beam polarization effects and the
high-energy behaviour of the cross section are also analyzed.

\end{abstract}

\newpage

\section{Introduction}
There are at present several experiments searching for neutrinoless double beta
decay ($\obb$) that are running, under construction or in the
planning phase \cite{Avignone:2007fu,GomezCadenas:2011it}. Observation 
of $\obb$ will be proof of lepton number violation, but 
extracting more specific information requires an assumption as to the underlying
mechanism of the process. While one
usually assumes that massive Majorana neutrinos will be the leading
contribution, there are many other particle 
physics candidates that can lead to $\obb$
\cite{Vergados:2002pv,Rodejohann:2011mu}. These include, to name a few, particles in $R$-parity violating
supersymmetric theories, heavy (including fourth generation) neutrinos,
leptoquarks, Majorons, as well as particles arising in
extra-dimensional and left-right symmetric theories. Indeed, current
limits on the lifetime of $\obb$ can be used to set 
constraints on a variety of particle physics parameters
\cite{Rodejohann:2011mu}.  In this paper we will focus on $\obb$
within left-right symmetric theories, and propose to test one of the
possible diagrams at a linear collider running in like-sign electron
mode. 

It is obvious that in comparison to $\obb$ a linear collider has the advantage of providing an extremely clean environment to test lepton number violation. While $\obb$, 
\be \label{eq:0bb}
(A,Z) \to (A,Z+2) + 2 e^- \, , 
\ee
is plagued by nuclear physics uncertainties, linear collider processes
such as 
\be\label{eq:inv}
e^- \, e^- \to W^- \, W^-
\ee
directly test the central part of most $\obb$-diagrams, see for
instance
Figs.~\ref{fig:fd_0nubb_LL_RR}, \ref{fig:triplet_R} and \ref{fig:fd_0nubb_lambda_eta}.
Indeed, the process (\ref{eq:inv}), often called inverse neutrinoless
double beta decay, has been proposed frequently in the past 
\cite{Rizzo:1982kn,London:1987nz,Huitu:1993gf,Rizzo:1994tk,Helde:1994xj,Gluza:1995ky,Belanger:1995nh,Ananthanarayan:1995cn,Rizzo:1995mr,Heusch:1996an,Gluza:1997kg,Duka:1998fv,Maalampi:1998pq,Rodejohann:2010jh,Kom:2011nc}
to test lepton number conservation and to check the mechanism of
$\obb$. Here we revisit the process in which left-handed and right-handed $W$-bosons of an $SU(2)_L \times SU(2)_R$ symmetric theory are produced \cite{London:1987nz,Helde:1994xj,Gluza:1995ky}: 
\be\label{eq:main}
e^- \, e^- \to W_L^- \, W_R^- \, ,  
\ee
depicted in Fig.~\ref{fig:fd_inv_0nubb}. 
The corresponding double beta diagram is the so-called $\lambda$-diagram, see Fig.~\ref{fig:obb_LR_lambda}. We will argue that from the
many possible $\obb$-diagrams in left-right symmetric theories, this
is the one which promises the largest cross section at an
electron-electron machine. We evaluate the
cross section and apply current limits from $\obb$ to it. Beam
polarization issues are also considered, and the high energy behaviour
of the cross section is analyzed. 

Note that lepton number violating processes can be tested at hadron colliders via the process $q\bar{q}'\rightarrow \ell\ell j j$, i.e. production of like-sign dileptons plus jets, which could proceed via the exchange of heavy neutrinos and right-handed $W$ bosons. Several studies in this direction have been performed \cite{Keung:1983uu,Ferrari:2000sp,Tello:2010am,Nemevsek:2011aa}, although the $\lambda$-diagram was not included\footnote{For other collider probes of $\obb$, see \cite{Allanach:2009iv,Allanach:2009xx}.}.
Recent analyses by the CMS~\cite{CMS_WR:2011} and ATLAS~\cite{Aad:2011yg,ATLAS_WR_NR:2012} collaborations using data from $pp$ collisions at $\sqrt{s}=7$ TeV have excluded certain regions in the $\mwr-M_R$ mass plane, where $M_R$ is the right-handed neutrino mass scale. The ATLAS limit on $\mwr$ extends up to nearly 2.5 TeV for some values of $M_R$, assuming that $\mwr > M_R$. These analyses also assume negligible left-right mixing between light and heavy neutrinos and between gauge bosons.

Other features of the left-right symmetric model include the existence of a new neutral gauge boson $Z'$, which could in principle be seen at both $pp$ \cite{Ferrari:2000sp} and $e^+e^-$ colliders \cite{Almeida:2004hj} (see Ref.~\cite{Langacker:2008yv} for a review and further references). Roughly speaking, since $m_{Z'} \simeq 1.7 \mwr$ (see Appendix~\ref{sec:LR}), the cross sections for processes such as $pp\rightarrow Z' \rightarrow \ell^+\ell^-$ will be lower than those involving charged gauge bosons. At linear colliders the $Z'$ could mediate new four fermion interactions, i.e. $e^+e^- \rightarrow Z' \rightarrow f\bar{f}$, and could be detected due to interference with the virtual $\gamma$ and $Z$ contributions. In addition, the model includes doubly charged scalar bosons, which could be produced in $pp$ and $e^+e^-$ collisions \cite{Huitu:1996su,Melfo:2011nx}; the latest ATLAS limits are $m^{\pm\pm}_{\delta_L} > 244$~GeV and $m^{\pm\pm}_{\delta_R} > 209$~GeV \cite{Aad:2012cg}. $s$-channel production of $\delta^{--}$ at like-sign linear colliders has been studied in, e.g. Refs.~\cite{Barenboim:1996pt,Rodejohann:2010bv}. 

The detection of a neutral gauge boson or doubly charged scalars would provide alternative tests of the left-right model, although the former has no immediate connection to $\obb$. Here we focus on the $\lambda$-diagram at an $e^-e^-$ machine, and show that it is observable if the $W_R$ mass and the life-time of $\obb$ are close to their current limits. Note that this process can be tested not only at a linear collider, but also due to its unique angular distribution in $\obb$~\cite{Arnold:2010tu}, and that it has not yet been studied at hadron colliders. Our analysis is thus complementary to those in Refs.~\cite{Keung:1983uu,Ferrari:2000sp,Tello:2010am,Nemevsek:2011aa}. \\

The paper is built up as follows: in Section \ref{sec:lr_0vbb} we summarize the various diagrams for $\obb$ within left-right symmetric theories, and argue that the so-called  $\lambda$-diagram looks most promising for tests at a linear collider. Then in Section \ref{sec:xs} we discuss the cross section, including the effects of beam polarization. Details of left-right symmetric theories, a study of the high
energy behaviour, and the helicity amplitudes of the process are delegated to the appendices, and we conclude in Section \ref{sec:concl}. 

\section{$\obb$ in left-right symmetric models} \label{sec:lr_0vbb}

\begin{figure}[t]
 \centering
 \subfigure[]{\label{fig:obb_LR_standard}
 \includegraphics[width=0.35\textwidth]{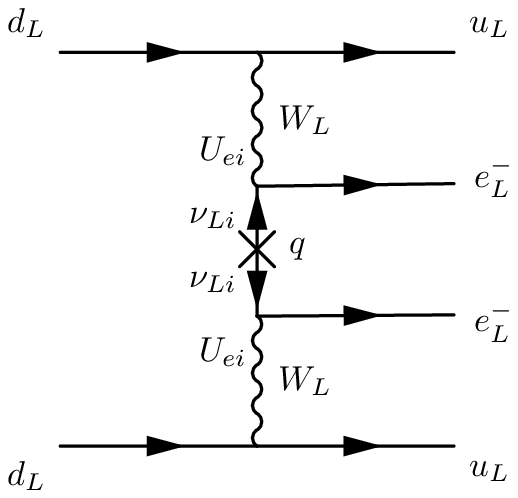}}
 \hspace{5mm}
 \subfigure[]{\label{fig:obb_LR_NR}
 \includegraphics[width=0.35\textwidth]{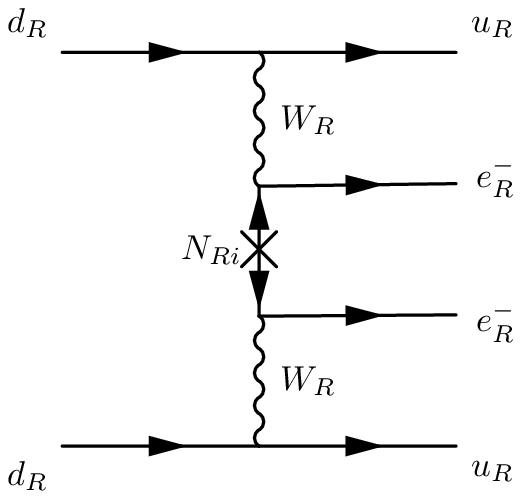}}
 \caption{Feynman diagrams of double beta decay in the left-right
symmetric model, mediated by (a) light neutrinos (the standard
mechanism) and by (b) heavy neutrinos in the presence of right-handed
currents. There is also a diagram with heavy neutrino exchange and
left-handed currents, as well light neutrino exchange and right-handed
currents, the latter is negligible.} 
 \label{fig:fd_0nubb_LL_RR}
\end{figure}

Here we summarize the various possible diagrams for $\obb$ in
left-right symmetric models (for one of the first analyses on this
topic, see \cite{Hirsch:1996qw}). 
Details of the theory are
delegated to Appendix \ref{sec:LR}; here it suffices to know that 
there are left- and right-handed currents with the associated gauge
bosons $W_L$ and $W_R$ (that can mix with each other), 
Higgs triplets $\Delta_L$ and $\Delta_R$ coupling to left- and
right-handed leptons, respectively, as well as light left-handed and heavy
right-handed Majorana neutrinos that can also mix with each other. 
With this particle content, one can construct the diagrams leading to $\obb$ displayed
in Figs.~\ref{fig:fd_0nubb_LL_RR}, \ref{fig:triplet_R} and \ref{fig:fd_0nubb_lambda_eta}. 
They can be categorized in terms of their topology and the helicity of the final state electrons. 
We will discuss them in detail; the limits on the particle physics
parameters are taken from Ref.~\cite{Rodejohann:2011mu}.  
 
\begin{itemize}
\item Fig.~\ref{fig:obb_LR_standard} is the standard diagram, whose
amplitude is proportional to 
\begin{equation}
 {\cal A}_{LL} \simeq G_F^2 \frac{\mee}{q^2}\, ,
\end{equation}
where $|q^2| \simeq (100$ MeV)$^2$ is the momentum exchange of the process. 
The particle physics parameter $\mee\equiv \left|\sum U_{ei}^2m_i\right|$ is called the effective mass,
and the suitably normalized dimensionless parameter describing lepton number violation is 
\begin{equation}
\eta_{LL} =  \frac{\mee}{m_e} = 
\frac{\left|\sum U_{ei}^2m_i\right|}{m_e} \ls 9.9 \times 10^{-7} \, .
\end{equation}
Here $U_{ei}$ is the (PMNS) mixing matrix of light neutrinos and $m_i$ are the light neutrino masses.
\item Fig.~\ref{fig:obb_LR_NR} is the exchange of right-handed
neutrinos with purely right-handed currents. The amplitude is
proportional to 
\begin{equation}
 {\cal A}_{N_R} \simeq G_F^2 \left(\frac{\mwl}{\mwr}\right)^4 \sum_i
\frac{{V^*_{ei}}^2}{M_i}\, ,
\end{equation}
where $\mwr$ ($\mwl$) is the mass of the right-handed $W_R$
(left-handed $W_L$), $M_i$ the mass of the heavy neutrinos and $V$ the right-handed
analogue of the PMNS matrix $U$. The dimensionless particle physics parameter is  
\be
\eta_{N_R} = m_p \left(\frac{\mwl}{\mwr}\right)^4 \left| \sum_i
\frac{{V^*_{ei}}^2}{M_i}\right| \ls 1.7 \times 10^{-8}\, .
\ee

There is also a diagram with left-handed currents in which right-handed neutrinos are
exchanged. The amplitude is proportional to 
\begin{equation}
 {\cal A}_{N^{\rm (LH)}_R} \simeq G_F^2  \sum_i
\frac{S_{ei}^2}{M_i}\, , 
\end{equation}
with $S$ describing the mixing of the heavy neutrinos with left-handed
currents. The limit is 
\be
\eta_{N^{\rm (LH)}_R} = m_p  \left| \sum_i\frac{S_{ei}^2}{M_i} \right| 
\ls 1.7 \times 10^{-8} \, . 
\ee
Another possible diagram is light neutrino exchange with right-handed currents, which is however highly suppressed.
\begin{figure}[t]
 \centering
 \subfigure[]{\label{fig:tripletRR}
 \includegraphics[width=0.35\textwidth]{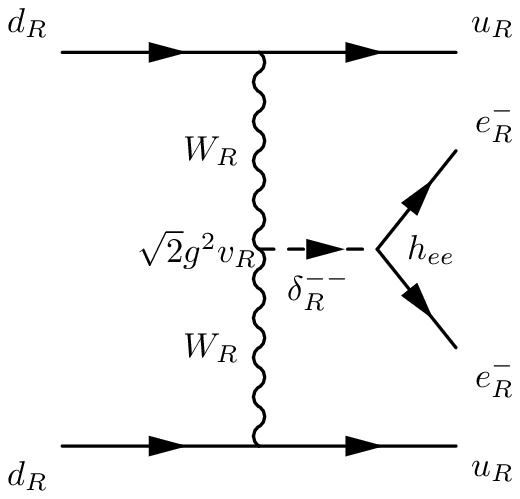}}
 \hspace{5mm}
 \subfigure[]{\label{fig:tripletLL}
 \includegraphics[width=0.35\textwidth]{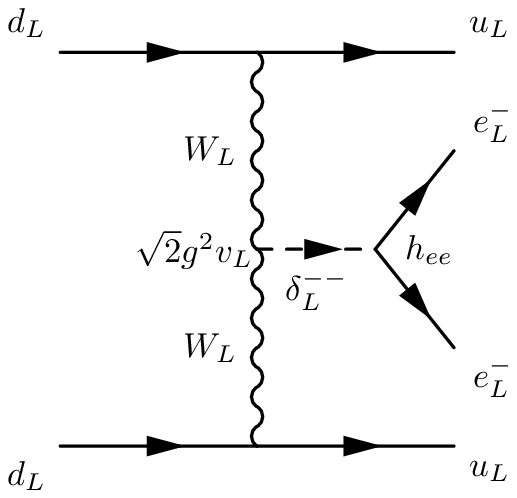}}
 \caption{Feynman diagrams of double beta decay in the left-right
symmetric model, mediated by doubly charged triplets:  (a) triplet of
$SU(2)_R$ and (b) triplet of $SU(2)_L$.} 
 \label{fig:triplet_R}
\end{figure}	

\item Fig.~\ref{fig:tripletRR} is a diagram with different topology,
mediated by the triplet of $SU(2)_R$. The amplitude is given by  
\begin{equation}
 {\cal A}_{\delta_R} \simeq G_F^2\left(\frac{\mwl}{\mwr}\right)^4\sum_i\frac{{V^*_{ei}}^2M_i}{m_{\delta^{--}_R}^2}\, ,
\end{equation}
and the dimensionless particle physics parameter is
\be 
\eta_{\delta_R}  = \frac{\left|\sum_i {V^*_{ei}}^2M_i\right|}{m_{\delta^{--}_R}^2\mwr^4}\frac{m_p}{G_F^2} \ls 6.9 \times 10^{-6}\, .
\ee
Here we have used the fact that the term $v_R h_{ee}$ is nothing but the $ee$ element of the right-handed Majorana neutrino mass matrix $M_R$ diagonalized by $V$, with $v_R$ the VEV of the triplet $\Delta_R$ and $h_{ee}$ the coupling of the triplet with right-handed
electrons.
\begin{figure}[t]
 \centering
 \subfigure[]{\label{fig:obb_LR_lambda}
 \includegraphics[width=0.35\textwidth]{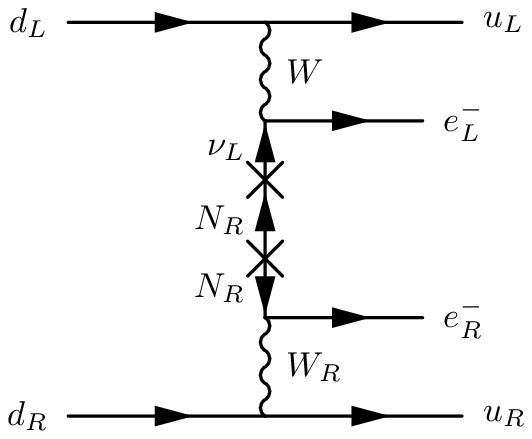}}
 \hspace{5mm}
 \subfigure[]{\label{fig:obb_LR_eta}
 \includegraphics[width=0.35\textwidth]{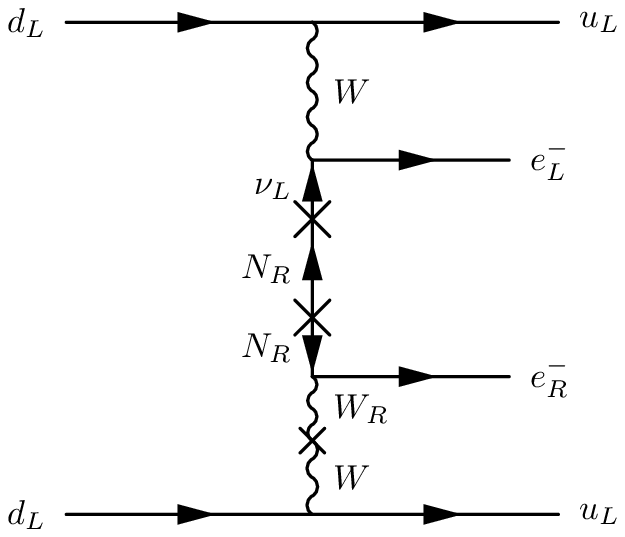}}
 \caption{Feynman diagrams of double beta decay in the left-right
symmetric model with final state electrons of different helicity: (a)
the $\lambda$-mechanism and (b) the $\eta$-mechanism due to gauge
boson mixing.} 
 \label{fig:fd_0nubb_lambda_eta}
\end{figure}
\item  Fig.~\ref{fig:tripletLL} is a diagram mediated by the triplet of $SU(2)_L$. 
The amplitude is given by 
\begin{equation}
 {\cal A}_{\delta_L} \simeq G_F^2 \frac{h_{ee} v_L}{m_{\delta^{--}_L}^2}\, .
\end{equation}
The diagram is suppressed with respect to the standard light neutrino
exchange by at least a factor $q^2/m_{\delta^{--}_L}^2$.
\item Fig.~\ref{fig:obb_LR_lambda} is a diagram in which the helicities of the final state electrons are different. It is called the $\lambda$-diagram, and has an amplitude reading 
\begin{equation}
 {\cal A}_\lambda \simeq G_F^2 \left(\frac{\mwl}{\mwr}\right)^2 \sum_i U_{ei} T^*_{ei}\, \frac{1}{q}\, ;
\end{equation}
the particle physics parameter is 
\be
\eta_{\lambda} = \left(\frac{\mwl}{\mwr}\right)^2\left|\sum_i U_{ei}T^*_{ei}\right| \ls 9 \times 10^{-7}\,,
\label{etal}
\ee
where $T^*_{ei}$ quantifies the mixing of light neutrinos with
right-handed currents [Eq.~\eqref{eq:Wmatrix}].
\item Finally, Fig.~\ref{fig:obb_LR_eta} is another diagram with mixed
helicity, possible due to $W_L-W_R$ mixing, which is described by the
parameter $\tan\zeta$ defined in Eq.~\eqref{eq:wlwr_mixing}. The
amplitude is  
\begin{equation}
 {\cal A}_\eta \simeq G_F^2\tan\zeta\sum_iU_{ei}T^*_{ei}\, \frac{1}{q}\, ,
\end{equation}
with particle physics parameter
\be
\eta_{\eta} = \tan\zeta\left|\sum_i U_{ei}T^*_{ei}\right| \ls 6 \times 10^{-9}\,.
\label{etaeta}
\ee
Note that in both the $\lambda$- and $\eta$-diagrams there are light
neutrinos exchanged (long-range diagrams), and the amplitude is proportional to the mixing matrix $T^*_{ei} = {\cal O}(M_D/M_R)$. One therefore needs both a non-zero $M_D$ and $M_R < \infty$, which is illustrated by the Dirac and Majorana mass terms in the propagator. In this case lepton number violation is implicit: the mixing $M_D/M_R$ vanishes for infinite Majorana mass. Ref.~\cite{Vergados:2002pv} gives a detailed explanation of how a
complicated cancellation of different nuclear physics amplitudes leads
to a limit on the $\eta$-diagram that is much stronger than the one on
the $\lambda$-diagram.  
\end{itemize}

Having written down all interesting diagrams, it is instructive to
discuss their expected relative magnitudes. For this naive exercise,
let us denote the masses of all particles belonging to the
right-handed sector ($M_i$, $W_R$ and $\delta^{--}_R$) as $R$. The matrices
$T$ and $S$ describing left-right mixing can be written as $L/R$, where $L$
is about $10^2$ GeV, corresponding to the weak scale, or the mass of
the $W_L$. The gauge boson mixing angle $\zeta$ is at most of order
$(L/R)^2$, and can be much smaller\footnote{See Eq.~\eqref{eq:zeta_def} and the comments just below it.}. The mixed $\lambda$- and $\eta$-diagrams in
Fig.~\ref{fig:fd_0nubb_lambda_eta} are of order $(L/R)^3 /q$, whereas
the purely right-handed short-range diagrams in
Figs.~\ref{fig:obb_LR_NR} (heavy neutrino exchange and right-handed
currents) and \ref{fig:tripletRR} ($SU(2)_R$ triplet exchange and
right-handed currents) are of order $L^4 /R^5$. Therefore, with $R$
being of order TeV, the mixed diagrams are expected to dominate
by a factor  $R^2/(L q) \sim 10^5$. In the same sense, the amplitudes 
of the mixed diagrams are also larger than the one for heavy neutrino
exchange with left-handed currents (which is proportional to
$L^2/R^3$).  Leaving these estimates aside, we continue with a purely
phenomenological analysis of the different diagrams at a linear collider.\\ 

\begin{figure}[t]
 \centering
 \subfigure[$t$-channel production]{\label{fig:wlwr_t}
 \includegraphics[width=0.35\textwidth]{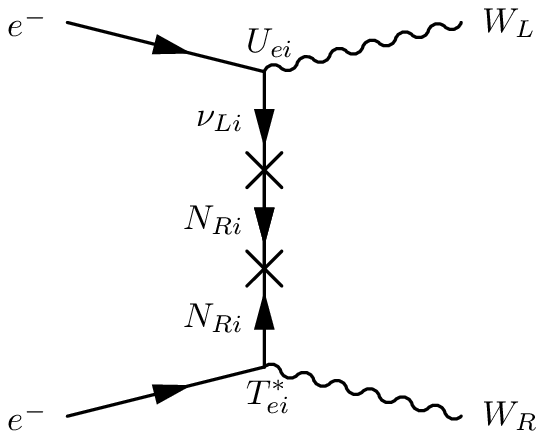}}
 \hspace{5mm}
 \subfigure[$u$-channel production]{\label{fig:wlwr_u}
 \includegraphics[width=0.35\textwidth]{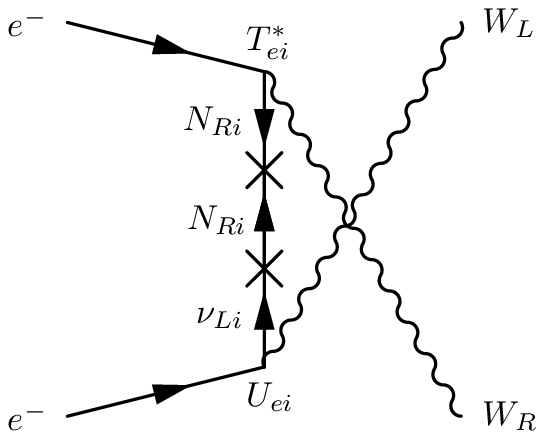}}
 \caption{Inverse neutrinoless double beta decay diagrams with $W_L$ and $W_R$ in the final state.}
 \label{fig:fd_inv_0nubb}
\end{figure}

For this exercise, let us use crossing symmetry to translate the $\obb$-diagrams from
Figs.~\ref{fig:fd_0nubb_LL_RR}--\ref{fig:fd_0nubb_lambda_eta} into
linear collider cross sections of the form $e^-e^-\rightarrow
W^-W^-$ (Fig.~\ref{fig:fd_inv_0nubb}). In each case the two gauge bosons can either have the same polarization ($W^-_LW^-_L$ or $W^-_RW^-_R$), in which case the process can be mediated by either Majorana neutrinos or Higgs triplets, or opposite polarizations ($W^-_LW^-_R$), only possible with the exchange of Majorana neutrinos plus non-zero left-right mixing. Since the limits on $W_R$ are 2.5 TeV
\cite{Maiezza:2010ic,Guadagnoli:2010sd}, diagrams with two $W_R$ are obviously disfavoured. 
In what regards diagrams with two $W_L$, both
the light and the heavy neutrino exchange can be shown to be
suppressed and unobservable, see Ref.~\cite{Rodejohann:2010jh} for a
recent reanalysis. The cross section corresponding to left-handed triplet exchange [Fig.~\ref{fig:tripletLL}] is proportional to $\sqrt{2}v_Lh_{ee} = [M_L]_{ee}$ (see Eq.~\eqref{eq:md_ml_mr} and Ref.~\cite{Rodejohann:2010bv}), so that it is suppressed by light neutrino mass. We are left with diagrams with only one $W_R$, i.e.~the mixed diagrams from Fig.~\ref{fig:fd_0nubb_lambda_eta}.  Noting
that the limit on the $\lambda$-diagram is less stringent by almost
two orders of magnitude with respect to the one for $\eta$ [compare Eqs.~\eqref{etal} and \eqref{etaeta}], 
we are led
to the conclusion that the $\lambda$-diagram is the most promising
channel to study.  Fig.~\ref{fig:fd_inv_0nubb} shows the relevant
Feynman diagram; its cross section will be evaluated in what follows. 
Let us note here that the SuperNEMO experiment has the possibility to
disentangle the $\lambda$-diagram from the standard one, because it
can probe the angular and energy correlation of the two emitted
electrons in $\obb$ \cite{Arnold:2010tu}. The mechanism is therefore
testable in a variety of ways.  

\section{\label{sec:xs}Cross section of $e^-e^-\rightarrow W^-_L W^-_R$}

\subsection{Cross section}

\begin{figure}
\begin{center}
\vspace{-1cm}
\includegraphics[angle=-90,width=0.7\textwidth]{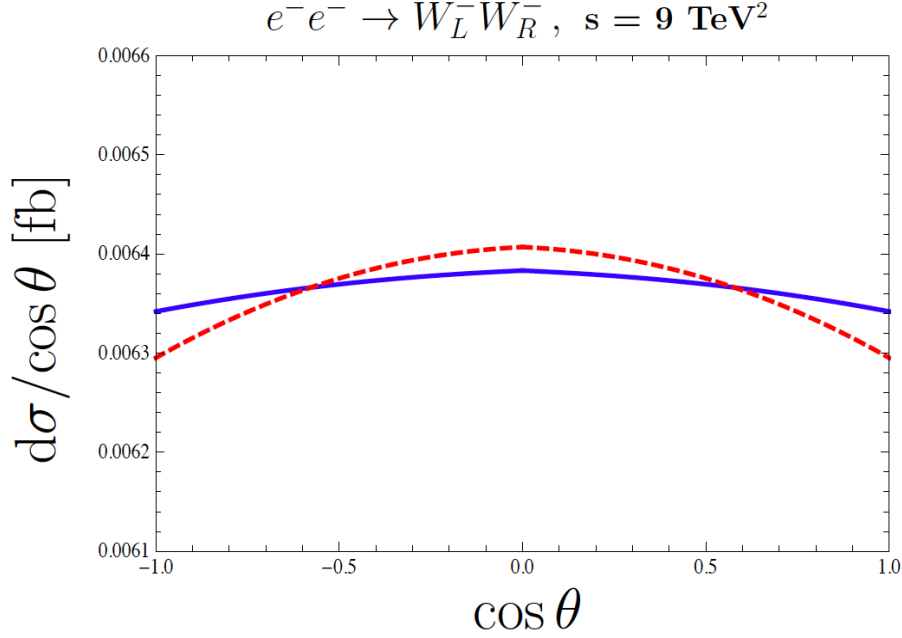} 
 \caption{Differential cross section for $e^-e^-\rightarrow W^-_L W^-_R$ with $\sqrt{s} = 3$~TeV and for both $\mwr = 2.5$~TeV (dashed red line) and 2.7~TeV (solid blue line), with the latter normalized to facilitate comparison.}
\label{fig:diff_cross_sect}
\end{center}
\end{figure}

The two possible channels for the process $e^-(p_1)\, e^-(p_2) \rightarrow W_L^-(k_1,\mu)\,  W_R^-(k_2,\nu)$ are shown in Fig.~\ref{fig:fd_inv_0nubb}. Here $p_{1,2}$ and $k_{1,2}$ are the momenta of the particles and $\mu$, $\nu$ the Lorentz indices of the $W$ polarization vectors. The matrix element is
\be 
-i {\cal M} = -i \left({\cal M}_t + {\cal M}_u  \right) , 
\ee
where the subscript denotes the $t$- or $u$-channel process.
In order to evaluate the differential cross section 
\be
\frac{d \sigma}{d \Omega} =  \frac{1}{64 \pi^2 \, s} 
   \frac 14 |\overline{\cal M} |^2 \, 
\sqrt{\frac{\lambda(s,\mwl^2,\mwr^2)}{\lambda(s,0,0)}} \, ,
\ee
where $\lambda(a,b,c) = a^2 + b^2 + c^2 - 2 \,(a\,b + a\,c +
b\,c)$, we need  
\be
|\overline{\cal M} |^2 = 
|\overline{{\cal M}_t} |^2 + |\overline{{\cal M}_u} |^2\,+  2{\rm Re} \left( 
\overline{ {\cal M}_t^\ast \, {\cal M}_u} 
\right) . 
\ee
The result is (neglecting $\mwl$)
\begin{align} \nonumber
|\overline{{\cal M}_t} |^2& = \frac{8\,G_F^2 \left|\sum_i U_{ei} \, T^*_{ei}\right|^2}{(t-m_{i}^2)^2} \left(\frac{\mwl}{\mwr}\right)^2\left\{ 4\mwl^4\mwr^2(t-\mwr^2) - t^2\left[t(s + t)-\mwr^2(2s + t)\right] \right.\\[1mm]  \nonumber
& + \left. \mwl^2 t\left[4 \mwr^4 + t(2s +t)- \mwr^2(4s + 5
t)\right]\right\}  , \\[1mm] 
|\overline{{\cal M}_u} |^2 & =|\overline{\cal M}_t |^2 \, (t \leftrightarrow u)\, , \label{eq:mat_elem}\\[1mm] 
\overline{ {\cal M}_t^\ast \, {\cal M}_u} & \propto {\rm Tr}\{P_R\gamma_\nu\ta{q}\, \, \, \gamma_\mu\ta{p}\, \, \, _1\gamma_\alpha\ta{\tilde{q}}\, \, \,\gamma_\beta\ta{p}\, \, \, _2P_L\} = 0 \nonumber\, \, .
\end{align}
The interference term vanishes, since the final state particles are distinguishable. Fig.~\ref{fig:diff_cross_sect} shows the
differential cross section $d\sigma/d\cos\theta$ as a function of $\cos\theta$, for $\mwr = 2.5$~TeV and $2.7$~TeV, normalized with respect to each other (the cross section for $\mwr = 2.7$~TeV is actually a factor of two smaller). $d\sigma/d\cos\theta$ is practically flat, and approaches a straight line as $\mwr$ increases.

It is interesting to study the high energy behaviour of the total cross section in the case of light neutrino exchange. In the limit that $\sqrt{s} \rightarrow \infty$, the cross section becomes
\be
\sigma(e^- e^- \ra W_L^- W_R^-) \simeq \dfrac{G_F^2\, \mwr^2}{24 \, \pi \,\mwl^2 } \, s\,\eta_\lambda^2 \leq \,  8.8\times10^{-5}\left(\dfrac{\mwr}{\rm TeV}\right)^2\left(\dfrac{\sqrt{s}}{\rm TeV}\right)^2\left(\dfrac{\eta_\lambda}{9\times10^{-7}}\right)^2 \rm fb\, , \label{eq:cross_sect_high_energy}
\ee
where the upper bound on $\eta_\lambda$ is given in Eq.~\eqref{etal} and we have neglected the mass of the light neutrinos $m_i$ in the propagator. The apparent violation of unitarity can be explained by taking the full theory into account, in which case the cross section will vanish when $\sqrt{s} \rightarrow \infty$ and unitarity is restored (see Appendix \ref{sec:HE} for details). 

There is also another diagram analogous to
Fig.~\ref{fig:fd_inv_0nubb}, with heavy neutrinos exchanged. The
structure of the matrix elements is the same, we need only to
interchange $m_{i}\leftrightarrow M_i$, $U_{ei} \leftrightarrow
V^*_{ei}$ and $T^*_{ei} \leftrightarrow S_{ei}$, where $M_i$ is the
mass of the heavy neutrinos and $V^*_{ei}$ and $S_{ei}$ are 3$\times$3
mixing matrices defined in Eq.~\eqref{eq:Wmatrix}. In this case the rate for
double beta decay will be suppressed with respect to the case of light
neutrino exchange in the $\lambda$-diagram.

\begin{figure}
\begin{center}
\includegraphics[width=12cm]{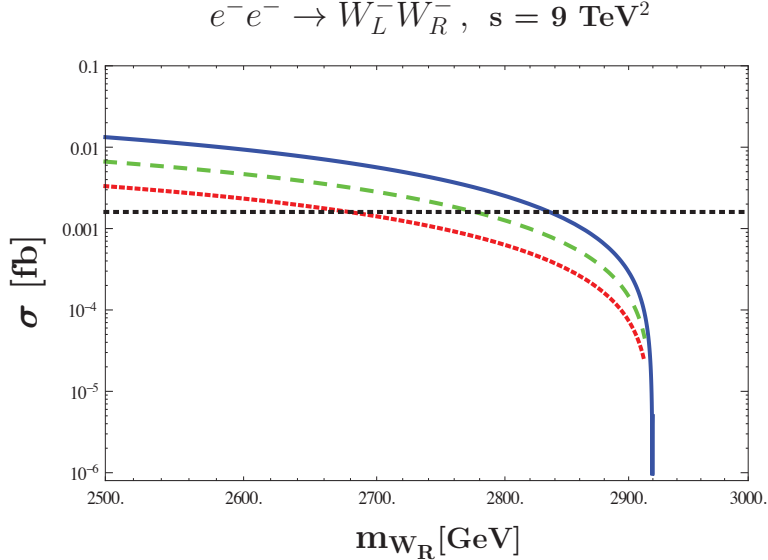} 
 \caption{Cross section for $e^-e^-\rightarrow W^-_L W^-_R$ with
$\sqrt{s} = 3$~TeV and three limits for the $\eta_\lambda$ parameter:
the solid (blue) line is for the current limit $\eta_\lambda = 9 \times
10^{-7}$, the dashed (green) line and the dotted (red) line are for
limits on  $\eta_\lambda$ improved by a factor $\sqrt{2}$ and $2$,
respectively. The dotted (black) horizontal line corresponds to the cross section
that would give five events at an integrated luminosity of 3000
fb$^{-1}$.} 
\label{fig:cross_sect}
\end{center}
\end{figure}

To calculate the total cross section the limits from $0\nu\beta\beta$
experiments as well as the allowed region for $\mwr$ must be taken
into account. Fig.~\ref{fig:cross_sect} shows the cross section for
$e^-e^-\rightarrow W^-_L W^-_R$ as a function of $\mwr$ for $\sqrt{s} =
3$ TeV, assuming only light neutrinos are exchanged and with three
different limits for $\eta_\lambda$: the solid (blue) line corresponds
to the present upper limit [Eq.~\eqref{etal}] given by $0\nu\beta\beta$ experiments, the dashed
(green) uses a limit improved by a factor of $\sqrt{2}$ and the dotted
(red) line is for a limit improved by a factor of 2. Note that a factor
of $x$ improvement in $\eta_\lambda$ corresponds to a factor of $x^2$
improvement in life-time.  
We also indicate the cross section that would give five events at an
integrated luminosity of 3000 $\rm fb^{-1}$ \cite{clic}, corresponding
to a few years of running. It is evident
that for $2.5\ {\rm TeV} \ls \mwr \ls 2.8\ {\rm TeV}$, enough events are
possible in case $0\nu\beta\beta$ is observed soon, and caused by the
$\lambda$-diagram. Note that since there is no Standard Model
background to the process, a small rate is tolerable. 

In the next subsection we
will show that polarization of the electron beams could be used to
enhance the cross section by up to a factor of two.  
Finally, we should note that in neutrinoless double beta decay
different contributions could interfere destructively. In this case
the bound on $\eta_\lambda$ would be relaxed and a larger cross
section is possible.

\subsection{\label{sec:pol}Polarized beams}
Future linear colliders have the possibility to polarize their
beams. In order to quantify the effects on our process, we define the
polarization for an electron beam $P_{1,2}$ as follows:  
\be
P_{1,2} \equiv \dfrac{N_R^{1,2}-N_L^{1,2}}{N_R^{1,2}+N_L^{1,2}}\, ,
\ee 
where $N_R$ and $N_L$ stand for the number of electrons having right- 
and left-handed helicity in the electron beam 1 or 2, respectively. If
beam 1 is fully left-handed, $P_1 = -1$, whereas for a fully
right-handed beam, $P_1 = +1$. 

\begin{table}[t]
\begin{center}
\begin{tabular}{|c|c||c|}
\hline
\multicolumn{2}{|c||}{beam polarization}&  $R$ \\ \cline{1-2}
1&2&\\ \hline \hline
0\%&0\%&1\\ \hline
90\% RH &0\% & $1$\\ \hline
50\% LH &50\% LH &$0.75$ \\ \hline
50\% LH &50\% RH &$1.25$ \\ \hline
80\% LH &50\% RH &$1.40$ \\ \hline
90\% LH &90\% RH & $1.81 $\\ \hline
90\% LH &80\% RH & $1.72 $\\ \hline
100\% LH &100\% RH & $2$ \\ \hline
\end{tabular}
\caption{Suppression or enhancement factors of the cross section with
polarized beams with respect to the unpolarized case.}\label{tab1} 
\end{center}
\end{table}

When the electron beam 1 has a polarization of $P_1$ and the electron
beam 2 has a polarization of $P_2$, the total cross section
$\sigma(P_1,P_2)$ of a process is calculated as
\bea\label{eq:pol}
\sigma(P_1,P_2) = \dfrac{1}{4}\left\{(1-P_1)(1+P_2)\sigma_{LR}\, + \,
(1-P_1)(1-P_2)\sigma_{LL}\,  \right. \\ \left. 
+\, (1+P_1)(1+P_2)\sigma_{RR}\, +\, (1+P_1)(1-P_2)\sigma_{RL}\right\} ,
\eea
where $\sigma_{LR}$ stands for the cross section of the process when
both electron beams are 100\% polarized, one left-handed and the other
right-handed; $\sigma_{RL}$, $\sigma_{LL}$ and
$\sigma_{RR}$ are defined in a similar way. In our case for the $\lambda$-diagram
$\sigma_{LL}=\sigma_{RR}=0$, and $\sigma_{LR}$ ($\sigma_{RL}$) is the
cross section that would arise from the $t$-channel ($u$-channel)
diagram only. Furthermore, $\sigma_{LR} = \sigma_{RL}$. Thus, 
equation \eqref{eq:pol} simply becomes 
\be
\sigma(P_1,P_2) = \sigma(P_2,P_1) = \dfrac{\sigma_{LR}}{2}\left(1 - P_1 \, P_2\right) . 
\ee
Table \ref{tab1} gives numerical examples. We have defined the ratio
$R$ between the cross section of polarized and unpolarized beams: 
\be
R \equiv \dfrac{\sigma(P_1,P_2)}{\sigma(0,0)} =  1 - P_1 \, P_2
 \, . 
\ee
Obviously, $\sigma(0,0)$ is the total cross section calculated
before. We see that the event numbers can in principle be doubled. 
Furthermore, polarization could be used as an additional method to
distinguish different mechanisms for processes of the form $e^- e^-
\to W^- W^-$. For instance, the process $e^- e^- \to 4 $
jets~\cite{Kom:2011nc} mediated by $R$-parity violating supersymmetry, 
involves slepton exchange, which couple mainly to left-handed
electrons.

\section{\label{sec:concl}Conclusion}
We have considered in this paper the process $e^- e^-
\to W^-_L W^-_R$ as a clean check of the so-called $\lambda$-diagram
as the leading contribution to neutrinoless double beta decay. We argued that among
the many possible diagrams for $0\nu\beta\beta$ that are possible in
left-right symmetric theories, it is the most promising one at a
linear collider. Indeed, it may be possible to observe the process at
a linear collider with center-of-mass energy of 3 TeV. It is however
necessary that both the mass of the $W_R$ and the life-time of
$0\nu\beta\beta$ are close to their current experimental limits. We
have also considered beam polarization effects and the high energy
behaviour of the total cross section, as well as the individual amplitudes.

\begin{center}
\large{\bf Acknowledgements}
\end{center}
We thank Steve Kom for helpful comments. 
This work was supported by the ERC under the Starting Grant MANITOP
(JB and WR) as well as by a CSIC JAE predoctoral fellowship, by the Spanish
MEC under grants FPA2008-00319/FPA, FPA2011-22975 and MULTI-DARK
CSD2009-00064 (Consolider-Ingenio 2010 Programme), by
Prometeo/2009/091 (Generalitat Valenciana) and by the EU ITN UNILHC
PITN-GA-2009-237920 (LD). LD would like to thank the Max-Planck-Institut
f\"{u}r Kernphysik in Heidelberg for kind hospitality. 

\csection{Appendix}
\appendix
\renewcommand{\theequation}{A-\arabic{equation}}
\setcounter{equation}{0}

\section{\label{sec:LR}Details of the left-right symmetric model}

In the left-right symmetric model \cite{Mohapatra:1974gc,Pati:1974yy,Senjanovic:1975rk,Mohapatra:1980yp,Deshpande:1990ip}, the Standard Model is extended to include the gauge group $SU(2)_R$ (with gauge coupling $g_R \neq g_L$), and right-handed fermions are grouped into doublets under this group. Thus we have the following fermion
particle content under $SU(2)_L \times SU(2)_R \times U(1)_{B-L}$: 
\begin{eqnarray}
 L'_{iL} &= \begin{pmatrix}\nu'_i \\ \ell'_i\end{pmatrix}_L \, \sim
(\mathbf{2},\mathbf{1},\mathbf{-1})\, , & L'_{iR} =
\begin{pmatrix}\nu'_i \\ \ell'_i\end{pmatrix}_R \sim
(\mathbf{1},\mathbf{2},\mathbf{-1}) \, , \\[1mm]
  Q'_{iL} &= \begin{pmatrix}u'_i \\ d'_i\end{pmatrix}_L \, \sim
(\mathbf{2},\mathbf{1},\mathbf{\frac{1}{3}})\, , & Q'_{iR} =
\begin{pmatrix}u'_i \\ d'_i\end{pmatrix}_R \sim
(\mathbf{1},\mathbf{2},\mathbf{\tfrac{1}{3}})\, , 
\end{eqnarray}
with the electric charge given by $Q=T_L^3+T_R^3+\frac{B-L}{2}$ and
$i=1,2,3$. The subscripts $L$ and $R$ are associated with the
projection $P_{L,R} = \frac 12 (1 \mp \gamma_5)$. 
In order to break the gauge symmetry and allow Majorana
mass terms for neutrinos one introduces the Higgs triplets 
\begin{equation}
 \Delta_{L,R} \equiv \begin{pmatrix} \delta_{L,R}^+/\sqrt{2} & \delta_{L,R}^{++} \\ \delta_{L,R}^0 & -\delta_{L,R}^+/\sqrt{2} \end{pmatrix},
\end{equation}
with $\Delta_L \sim (\mathbf{3},\mathbf{1},\mathbf{2})$ and $\Delta_R
\sim (\mathbf{1},\mathbf{3},\mathbf{2})$; the electroweak symmetry is
broken by the bi-doublet scalar 
\begin{equation}
 \phi \equiv \begin{pmatrix} \phi_1^0 & \phi_2^+ \\ \phi_1^- & \phi_2^0 \end{pmatrix} \sim (\mathbf{2},\mathbf{2},\mathbf{0})\, .
\end{equation}
The relevant Lagrangian in the lepton sector is
\begin{align}
 {\cal L}^{\ell}_{Y} = &-\overline{L}'_{L}(f\phi +
g\tilde{\phi})L'_{R} -\overline{L}'^c_{L}i\sigma_2\Delta_L h_L L'_{L} -
\overline{L}'^c_{R}i\sigma_2\Delta_R h_R L'_{R} + {\rm h.c.}, 
\end{align}
where $\tilde{\phi} \equiv \sigma_2\phi^*\sigma_2$; $f, g$ and $h_{L,R}$ are
matrices of Yukawa couplings and charge conjugation is defined as 
\begin{equation}
 \psi_{L,R}^c \equiv {\cal C}\overline{\psi}_{L,R}^T\,, 
\quad {\cal C} \equiv i\gamma_0\gamma_2\, . 
\end{equation}
If one assumes a discrete LR symmetry in addition to the additional gauge symmetry, the gauge couplings become equal ($g_L=g_R=g$) and one obtains relations between the Yukawa coupling matrices in the model. With a discrete parity symmetry it follows that $h_L=h_R^*$, $f=f^\dagger$, $g=g^\dagger$; with a charge conjugation symmetry $h\equiv h_L=h_R$, $f=f^T$, $g=g^T$. 

Making use of the gauge symmetry to eliminate complex phases, the most
general vacuum is 
\begin{gather} 
 \langle \phi\rangle = \begin{pmatrix} \kappa_1/\sqrt{2} & 0 \\ 0 &
\kappa_2e^{i\alpha}/\sqrt{2} \end{pmatrix}, \quad \langle \Delta_{L}
\rangle = \begin{pmatrix} 0 & 0 \\ v_{L}e^{i\theta_L}/\sqrt{2} & 0
\end{pmatrix}, \quad \langle \Delta_{R} \rangle = \begin{pmatrix} 0 &
0 \\ v_{R}/\sqrt{2} & 0 \end{pmatrix}. 
\end{gather}
After spontaneous symmetry breaking, the mass term for the charged
leptons is 
\begin{equation}
 {\cal L}^{\ell}_{\rm mass} = -\overline{\ell}'_L M_\ell \ell'_R + {\rm h.c.},
\end{equation}
where the mass matrix
\begin{equation}
 M_\ell = \frac{1}{\sqrt{2}}(\kappa_2e^{i\alpha}f + \kappa_1g) \neq M^\dagger_\ell
\end{equation}
can be diagonalized by the bi-unitary transformation
\begin{equation}
 \ell'_{L,R} \equiv V_{L,R}^{\ell}\ell_{L,R}\, , \quad V_L^{\ell\dagger}M_\ell V_R^{\ell} = {\rm diag}(m_e,m_\mu,m_\tau)\, .
\end{equation}
In the neutrino sector we have a type I + II seesaw scenario,
\begin{align}
 {\cal L}^\nu_{\rm mass} = -\tfrac{1}{2}\overline{n'_L} M_\nu n'^c_L + {\rm h.c.} 
 = -\tfrac{1}{2}\begin{pmatrix} \overline{\nu'_L} \
\overline{{\nu'_R}^c}\end{pmatrix} \begin{pmatrix} M_L & M_D \\ M_D^T &
 M_R\end{pmatrix} \begin{pmatrix} {\nu'_L}^c \\ \nu'_R\end{pmatrix} + {\rm h.c.}\,, 
\label{eq:lag_nu_mass}
\end{align}
with
\begin{gather}
 M_D = \frac{1}{\sqrt{2}}(\kappa_1 f + \kappa_2e^{-i\alpha}g)\, ,
\quad M_L = \sqrt{2}v_Le^{i\theta_L}h\, , \quad M_R = \sqrt{2}v_R h\,
. \label{eq:md_ml_mr}
\end{gather}
Assuming that $M_L \ll M_D \ll M_R$, the light neutrino mass matrix
can be written in terms of the model parameters as 
\begin{align}
 m_\nu = M_L - M_DM_R^{-1}M_D^T
 = \sqrt{2}v_Le^{i\theta_L}h - \frac{\kappa_+^2}{\sqrt{2}v_R}h_Dh^{-1}h_D^T\, ,
\end{align}
where
\begin{equation}
 h_D \equiv \frac{1}{\sqrt{2}}\frac{\kappa_1f +
\kappa_2e^{-i\alpha}g}{\kappa_+}\,, \quad  
 \kappa_+^2 \equiv |\kappa_1|^2+|\kappa_2|^2\, .
\end{equation}
The symmetric $6\times6$ neutrino mass matrix $M_\nu$ in
Eq.~(\ref{eq:lag_nu_mass}) is diagonalized by the unitary $6\times6$
matrix \cite{Schechter:1981cv,Grimus:2000vj,Hettmansperger:2011bt} 
\begin{equation}\label{eq:Wmatrix}
 W \equiv \begin{pmatrix} V_L^{\nu} \\ V_R^{\nu} \end{pmatrix} =
\begin{pmatrix} U & S \\ T & V\end{pmatrix} \simeq \begin{pmatrix} 1_{3\times 3} &
M_DM_R^{-1} \\ -{M_R^{-1}}^*M_D^\dagger & 1_{3\times 3}\end{pmatrix}
\begin{pmatrix} U_{\rm PMNS} & 0 \\ 0 & V_R \end{pmatrix}\, 
\end{equation}
to $W^\dagger M_\nu W^* = {\rm diag}(m_1,m_2,m_3,M_1,M_2,M_3)$, where
the matrices $U_{\rm PMNS}$ and $V_R$ are defined by 
\bea
 M_L - M_DM_R^{-1}M_D^T = U_{\rm PMNS}\,{\rm diag}(m_1,m_2,m_3)U_{\rm
PMNS}^T \, , \\ 
M_R = V_R\,{\rm diag}(M_1,M_2,M_3)V_R^T\, . 
\eea
The neutrino mass eigenstates $n = n_L + n^c_L=n^c$ are defined by
\begin{align}
 n'_L &= \begin{pmatrix} \nu'_L \\ {\nu'_R}^c \end{pmatrix} = W n_L =
\begin{pmatrix} U & S \\ T & V \end{pmatrix}\begin{pmatrix} \nu_L \\
N^c_R \end{pmatrix} \\[1mm] 
 n'^c_L &= \begin{pmatrix} {\nu'_L}^c \\ \nu'_R \end{pmatrix} =
W^*n^c_L = \begin{pmatrix} U^* & S^* \\ T^* & V^*
\end{pmatrix}\begin{pmatrix} \nu^c_L \\ N_R \end{pmatrix} . 
\end{align}
Note that the unitarity of $W$ leads to the useful relations
\begin{equation}\label{eq:unit}
V_L^\nu V_L^{\nu\dagger} = UU^\dagger + SS^\dagger = \mathbb{1} =
V_R^\nu V_R^{\nu\dagger}=TT^\dagger+VV^\dagger\, \quad {\rm and} \quad
V_L^\nu V_R^{\nu \dagger} = UT^\dagger+SV^\dagger=0\, ,
\end{equation}
with the unitary $3\times6$ matrices $V^\nu_L = (U \ \ S)$ and $V^\nu_R = (T \ \ V)$ defined in Eq.~\eqref{eq:Wmatrix}. 

The leptonic charged current interaction in the flavour basis is
\bea
 {\cal L}^{\rm lep}_{CC} =
\frac{g}{\sqrt{2}}\left[\overline{\ell'}\gamma^\mu(P_L \cos\zeta -
P_R\sin\zeta\, e^{-i\alpha})\nu' W_{1\mu}^- \right. \\[1mm] 
+ \left. \overline{\ell'}\gamma^\mu(P_L \sin\zeta\, e^{i\alpha}  +
P_R\cos\zeta )\nu' W_{2\mu}^-\right] + {\rm h.c.}, 
\eea
where
\begin{equation}
 \begin{pmatrix} W_L^{\pm} \\ W_R^{\pm} \end{pmatrix} =
\begin{pmatrix} \cos\zeta & \sin\zeta\, e^{i\alpha} \\ -\sin\zeta\,
e^{-i\alpha} & \cos\zeta \end{pmatrix} \begin{pmatrix} W_1^{\pm} \\
W_2^{\pm}\end{pmatrix}
\label{eq:wlwr_mixing}
\end{equation}
characterizes the mixing between left- and right-handed gauge
bosons, with $\tan2\zeta = -\frac{2\kappa_1\kappa_2}{v_R^2-v_L^2}$. With negligible mixing the gauge boson masses become
\begin{equation}
 \mwl \simeq m_{W_1} \simeq \frac{g}{2}\kappa_+\, , \quad {\rm and} \quad \mwr \simeq m_{W_2} \simeq \frac{g}{\sqrt{2}}v_R\, , \label{eq:mw1_2}
\end{equation}
and assuming that\footnote{This is justified if one assumes no cancellations in generating quark masses \cite{Zhang:2007da}.} $\kappa_2<\kappa_1$, it follows that
\begin{equation}
 \zeta \simeq -\kappa_1\kappa_2/v_R^2 \simeq -2\frac{\kappa_2}{\kappa_1}\left(\frac{\mwl}{\mwr}\right)^2,
\label{eq:zeta_def}
\end{equation}
so that the mixing angle $\zeta$ is at most\footnote{Although the experimental limit is $\zeta < 10^{-2}$ \cite{Nakamura:2010zzi}, for $\mwr = {\cal O}({\rm TeV})$ one has $\zeta \ls 10^{-3}$ \cite{Langacker:1989xa}; supernova bounds for right-handed neutrinos lighter than 1 MeV are even more stringent ($\zeta < 3 \times 10^{-5}$) \cite{Raffelt:1987yt,Langacker:1989xa,Barbieri:1988av}.} the square of the ratio of left and right scales $(L/R)^2$.
The charged current then becomes 
\bea 
 {\cal L}^{\rm lep}_{CC} =
\frac{g}{\sqrt{2}}\left[\overline{\ell'_L}\gamma^\mu \nu'_L W_{L\mu}^-
+ \overline{\ell'_R}\gamma^\mu \nu'_R W_{R\mu}^-\right] + {\rm h.c.}
\\[1mm] 
 = \frac{g}{\sqrt{2}}\left[\overline{\ell_L}\gamma^\mu K_L n_L
W_{L\mu}^- + \overline{\ell_R}\gamma^\mu  K_R n^c_L W_{R\mu}^-\right]
+ {\rm h.c.} 
\label{eq:lag_cc_lr}
\eea
Here $K_L$ and $K_R$ are $3\times6$ mixing matrices
\begin{equation}
 K_{L} \equiv V_L^{\ell\dagger}V_L^{\nu}\, , \quad {\rm and} \quad
K_{R} \equiv V_R^{\ell\dagger}V_R^{\nu*}\, , 
\end{equation}
connecting the three charged lepton mass eigenstates $\ell_i$ to the six
neutrino mass eigenstates $(\nu_i,N_i)^T$, ($i=1,2,3$), with [using Eq.~(\ref{eq:unit})]
$K_L K_L^\dagger = K_RK_R^\dagger = \mathbb{1}$ and $K_LK_R^T = 0$. 

Note that in this model one also expects a new neutral gauge boson, $Z'$, which mixes with the standard model $Z$ boson. The mass eigenstates $Z_{1,2}$ have the masses
\begin{equation}
 m_{Z_1} \simeq \frac{g}{2\cos\theta_W}\kappa_+\, , \quad {\rm and} \quad m_{Z_2} \simeq \frac{g\cos\theta_W}{\sqrt{\cos2\theta_W}}v_R\, ,
\label{eq:mz1_2}
\end{equation}
where $g = e/\sin\theta_W$ and the $U(1)$ coupling constant is $g'\equiv e/\sqrt{\cos2\theta_W}$. Again one expects the mixing to be of order $(L/R)^2$. Eqs.~\eqref{eq:mw1_2} and \eqref{eq:mz1_2} imply that $m_{Z_2} \simeq 1.7 m_{W_2}$.

\section{\label{sec:HE}High energy behaviour of $e^-e^-\rightarrow W^-_LW^-_R$}

Naively, the high-energy limit of the cross section is obtained by
neglecting the neutrino mass in the propagator [see Eq.~\eqref{eq:mat_elem}], i.e.
\be
\sigma \propto  \left(\sum_{i} U_{ei} T^*_{ei} \right)^2\,  ,
\ee
which does not seem to vanish. However, one needs to consider the full
theory. In calculating the cross section one combines two terms from
the Lagrangian in 
Eq.~\eqref{eq:lag_cc_lr}: 
\begin{equation}
 \sum_i\left[\overline{e}\,\gamma^{\mu} (K_L)_{ei}P_L n_{i}
W^-_{L\mu}\right]\left[\overline{e}\,\gamma^{\nu} (K_R)_{ei}P_R n_{i}
W^-_{R\nu}\right] .
\end{equation}
The identity $\overline{e}\,\gamma^{\nu}P_Rn_i = -\overline{n^c_{i}}\,\gamma^\nu P_L e^c$ allows one to contract $n_{i}\overline{n^c_{i}}$ to a propagator, so that in the high energy limit the amplitude is proportional to 
\begin{equation}\label{unitarity}
 \sum_i (K_L)_{ei} (K_R)_{ei} = [K_LK_R^T]_{ee}\, ,
\end{equation}
instead of $\sum_i U_{ei} T^*_{ei}$ as in the naive case. As shown in
the previous subsection, $K_L K_R^T = 0$, which means that the cross
section vanishes in the high energy limit and unitarity is ensured. 

\section{\label{sec:HA}Helicity amplitudes for $e^-e^-\rightarrow
W^-_L W^-_R$}
It is an illustrative exercise to evaluate the helicity amplitudes of
the process $e^- e^- \to W_L^- W_R^-$, with the helicity of the
electrons and the polarization of the $W$-bosons fixed. Denoting electron ($W$-boson) momenta with $p_i$ ($k_i$), ($i=1,2$), the process is
\begin{equation}
 e^-(p_1,\lambda_1) \,   e^-(p_2,\lambda_2) \to W_L^-(k_1,\tau_1) \,  W_R^-(k_2,\tau_2)\, ,
\end{equation}
where $\lambda_{1,2} = \pm \frac{1}{2}$ and $\tau_{1,2} = 0,\pm 1$. Without loss of generality, one can choose $p_1$ and $p_2$ to be in the $\pm z$-directions, and assume that the final state particles propagate in the $x$--$z$ plane. The momenta are then given by
\begin{equation}
 p_{1,2}^\mu = \left(E,0,0,\pm E\right), \quad k_{1,2}^\mu = \left(E_{1,2},\pm|k|\vec{\mathbf{n}}\right),
\end{equation}
where $\vec{\mathbf{n}} = (\sin\theta,0,\cos\theta)$ and
\begin{equation}
 E = \frac{\sqrt{s}}{2}\, , \quad E_{1,2} = \frac{s\pm \mwl^2 \mp \mwr^2}{2\sqrt{s}}\, , \quad |k| = \frac{\sqrt{\lambda(s,\mwl^2,\mwr^2)}}{2\sqrt{s}}\, .
\end{equation}
The gauge boson polarization vectors can be defined by
\begin{align}
 \epsilon_{\tau_{1,2}=0}(k_1,k_2) &= \pm\frac{1}{m_{W_{L,R}}}(\pm |k|,E_{1,2}\sin\theta,0,E_{1,2}\cos\theta)\, , \\
 \epsilon_{\tau_{1,2}=\pm 1}(k_1,k_2) &= \frac{1}{\sqrt{2}}(0,\mp\tau_{1,2}\cos\theta,-i,\pm\tau_{1,2}\sin\theta)\, .
\end{align}
The helicity amplitudes are calculated from
\begin{align}
 {\cal M}_{\lambda_1\lambda_2\tau_1\tau_2} &= \frac{g^2}{2(t-m_i^2)}\bar{u}(p_1,\lambda_1)\gamma_\mu\slashed{q}\gamma_\nu P_L v(p_2,\lambda_2)\epsilon^{\mu *}(k_1,\tau_1)\epsilon^{\nu *}(k_2,\tau_2) \notag \\ 
 &+ \frac{g^2}{2(u-m_i^2)}\bar{u}(p_1,\lambda_1)\gamma_\mu\slashed{\tilde{q}}\gamma_\nu P_R v(p_2,\lambda_2)\epsilon^{\mu *}(k_2,\tau_2)\epsilon^{\nu *}(k_1,\tau_1)\, ,
\label{eq:gen_helicity_amp}
\end{align}
resulting in
\begin{align}
 {\cal M}_{\lambda\lambda00} &= -\lambda g^2\frac{\sin \theta  \left\{\sqrt{\lambda(s,\mwl^2,\mwr^2)}
   \left(s+\mwl^2+\mwr^2\right)-2\lambda\cos \theta  \left[\left(\mwl^2-\mwr^2\right)^2-s^2\right]\right\}}{4
   \mwl \mwr \left(q^2-m_i^2\right)}\, , \\[1mm]
 {\cal M}_{\lambda\lambda0\tau} &= \lambda g^2\sqrt{s} \left[(1+2\lambda\tau)\cos^2\frac{\theta}{2}+(1-2\lambda\tau)\sin^2\frac{\theta}{2}\right] \notag \\[1mm]
 & \quad \times \frac{\cos\theta \left(s+\mwl^2-\mwr^2\right)-4\lambda\tau\mwl^2+2\lambda\sqrt{\lamfunc}}{2\sqrt{2}\mwl
   \left(q^2-m_i^2\right)}\, , \\[1mm]
 {\cal M}_{\lambda\lambda\tau0} &= -\lambda g^2\sqrt{s} \left[(1-2\lambda\tau)\cos^2\frac{\theta}{2}+(1+2\lambda\tau)\sin^2\frac{\theta}{2}\right] \notag \\[1mm]
 & \quad \times \frac{\cos\theta \left(s-\mwl^2+\mwr^2\right)+4\lambda\tau\mwl^2+2\lambda\sqrt{\lamfunc}}{2\sqrt{2}\mwl
   \left(q^2-m_i^2\right)}\, , \\[1mm]
 {\cal M}_{\lambda\lambda\tau\tau} &= g^2\frac{\sin\theta \left[2\lambda\sqrt{\lamfunc}-2\lambda\tau\left(\mwl^2-\mwr^2\right)+ s \cos\theta\right]}{4\left(q^2-m_i^2\right)}\, , \\[1mm]
 {\cal M}_{\lambda\lambda\tau-\tau} &= -\frac{g^2s\sin\theta(\cos\theta-2\lambda\tau)}{4\left(q^2-m_i^2\right)}\, , \\[1mm]
 {\cal M}_{\lambda-\lambda 00} &= {\cal M}_{\lambda-\lambda0\tau} = {\cal M}_{\lambda-\lambda\tau0} = {\cal M}_{\lambda-\lambda\tau\tau} = {\cal M}_{\lambda-\lambda\tau-\tau} = 0\, ,
\end{align}
where $\lambda = \pm \frac{1}{2}$ and $\tau = \pm 1$, and $q^2 = t(u)$ when $\lambda = -\frac{1}{2}\left(+\frac{1}{2}\right)$. The amplitude vanishes whenever $\lambda_1 = -\lambda_2$, or in other words, when the two electrons have the same spin (note that one electron is described by a $v$ spinor in Eq.~\eqref{eq:gen_helicity_amp}, which means that its actual helicity is the opposite of the spinor's helicity). The amplitude is only non-zero when the electrons have opposite spin ($\lambda_1 = \lambda_2$); squaring and summing over boson polarizations gives the polarized cross sections $\sigma_{LR}$ and $\sigma_{RL}$ in Eq.~\eqref{eq:pol}, which correspond to the $t$- and $u$-channels respectively.

It is interesting to study the high energy behaviour of these helicity amplitudes. Explicitly, in the limit $\sqrt{s} \rightarrow \infty$ and neglecting neutrino mass one gets
\begin{align}
 {\cal M}_{\lambda\lambda00} &\xrightarrow{\sqrt{s} \rightarrow \infty} -\lambda\frac{g^2\sin\theta \,s}{2\mwl\mwr} \, ,\notag \\[1mm]
 {\cal M}_{\lambda\lambda0\tau} &\xrightarrow{\sqrt{s} \rightarrow \infty} -\frac{g^2\sqrt{s}\left[(1+2\lambda\tau)\cos^2\frac{\theta}{2}+(1-2\lambda\tau)\sin^2\frac{\theta}{2}\right]}{2\sqrt{2}\mwl}\, ,\notag \\[1mm]
 {\cal M}_{\lambda\lambda\tau0} &\xrightarrow{\sqrt{s} \rightarrow \infty} \frac{g^2\sqrt{s}\left[(1-2\lambda\tau)\cos^2\frac{\theta}{2}+(1+2\lambda\tau)\sin^2\frac{\theta}{2}\right]}{2\sqrt{2}\mwl} \, ,\\[1mm]
 {\cal M}_{\lambda\lambda\tau\tau} &\xrightarrow{\sqrt{s} \rightarrow \infty} -\lambda g^2\sin\theta \notag \, ,\\[1mm]
 {\cal M}_{\lambda\lambda\tau-\tau} &\xrightarrow{\sqrt{s} \rightarrow \infty} -\lambda\tau \frac{g^2\sin\theta(1-2\lambda\tau\cos\theta)}{1+2\lambda\cos\theta}\, . \notag
\end{align}
The amplitudes that contain at least one longitudinally polarized
$W$-boson ($\tau_{1,2}=0$) are divergent, whereas those with only
transverse polarizations ($\tau_{1,2}=\pm1$) are finite. Summing over
fermion spins and boson polarizations gives the result in
Eq.~\eqref{eq:cross_sect_high_energy}, and proper consideration of the
full theory will lead to a well-behaved total amplitude, in analogy to
Appendix~\ref{sec:HE}. 

\bibliographystyle{h-physrev}
\bibliography{lrbib.bib}

\end{document}